\documentclass[preprint,12pt]{elsarticle}

\usepackage{graphicx}
\usepackage{amsmath,amssymb}
\usepackage{amsthm}

\newcommand{\dv}{\,\mathrm{div}\,}
\newcommand{\rot}{\mathrm{rot}\,}
\newcommand{\const}{\mathrm{const}}
\newcommand{\bu}{\mathbf{u}}
\newcommand{\be}{\mathbf{e}}
\newcommand{\bx}{\mathbf{x}}

\newcommand{\bgamma}{\mbox{\boldmath $\gamma$}}

\newcommand{\balpha}{\mbox{\boldmath $\alpha$}}

\newcommand{\Dt}{D_t\,}
\newcommand{\Ds}{D_s\,}

\newcommand{\bb}{\mathbf{b}}

\newcommand{\bxi}{\mbox{\boldmath $\xi$}}
\newcommand{\bsigma}{\mbox{\boldmath $\sigma$}}
\newcommand{\btau}{\mbox{\boldmath $\tau$}}

\newcommand{\bbeta}{\mbox{\boldmath $\eta$}}

\newcommand{\bV}{\mathbf{V}}

\newcommand{\bB}{\mathbf{B}}

\newcommand{\pd}[2]{\frac{\partial #1}{\partial #2}}
\newcommand{\opd}[2]{\frac{\mathrm{d} #1}{\mathrm{d} #2}}
\newcommand{\pdd}[3]{\frac{\partial^2 #1}{\partial #2 \partial #3}}
\newcommand{\ppd}[2]{\frac{\partial^2 #1}{{\partial #2}^2 }}

\theoremstyle{plain}
\newtheorem{theorem}{Theorem}

\theoremstyle{definition}

\theoremstyle{remark}

\journal{Physical Letters A}

\begin{document}
\graphicspath{{C:/Sergey/Papers/2010/Pictures/}}
\begin{frontmatter}


\ead{sergey@hydro.nsc.ru}

\title{NATURAL CURVILINEAR COORDINATES FOR IDEAL MHD EQUATIONS. SOLUTIONS WITH CONSTANT TOTAL PRESSURE}


\author[a,b]{Sergey V. Golovin}

\address[a]{Lavrentyev Institute of Hydrodynamics SB RAS, 630090 Novosibirsk, Russia}
\address[b]{Department of Mechanics and Mathematics, Novosibirsk State University, 630090 Novosibirsk, Russia}

\begin{abstract}
Equations of magneto-gasdynamics in the natural curvilinear system of coordinates where trajectories and magnetic lines play a role of coordinate curves are reduced to the nonlinear vector wave equation coupled with the incompressibility condition in the form of the generalized Cauchy integral. The symmetry group of obtained equation, equivalence transformation, and group classification with respect to the constitutive equation are calculated. New exact solutions with functional arbitrariness describing non-stationary incompressible flows with constant total pressure are given by explicit formulae. The corresponding magnetic surfaces have the shape of deformed nested cylinders, tori, or knotted tubes.
\end{abstract}


\end{frontmatter}



\section*{Introduction}
%Mathematical model of ideal magneto-hydrodynamics (MHD) is widely applicable in astrophysical problems \cite{Somov2006}, in modeling of magnetic traps for plasma confinement \cite{Freidberg1987} and plasma acceleration devices \cite{Brushlinskij2009}, MHD-dynamo theories \cite{Moffatt1978en} etc. From the mathematical point of view the MHD model is attractive as a next step of complexity in comparison with the gas dynamics equations. The model describes evolution and mutual influence of velocity and magnetic vector fields. The magnetic field is carried by the fluid flow, which implies conservation of the topological structure of the magnetic lines. This generates interest to topological invariants of the flow \cite{Moffatt1992}, topological restrictions to the kinetic energy dissipation and related problems \cite{Berger2009}. The main complexity in the description of MHD flows is their essential multidimensionality, that is to say, even locally MHD is two-dimensional in comparison to the locally one-dimensional gas dynamics. Natural reductions of the dimension, such as one-dimensional flows with planar, cylindrical or spherical waves happened to be too restrictive from the physical point of view (e.g., Cowling anti-dynamo theorems \cite{Cowling1976}). At that, numerical simulations of 3D MHD flows is complicated by the presence of various types of strong and weak discontinuities \cite{JeffreyTaniuti1964, Kulikovskijetal2001}.

The topological structure of the magnetic field plays a significant role in applied problems (magnetic traps for plasma confinement, MHD-dynamo, etc.) and related mathematical theories \cite{Moffatt1978en,ArnoldKhesin1998, Berger2009}. Most of these theories require the magnetic field to be determined in a finite area of the 3D space, and satisfy natural conditions over the boundary of the area. However, the set of known examples of exact solutions of magneto-hydrodynamics equations (MHD) that describe such topologically nontrivial configurations is very poor. The main problem is that the existing methods (e.g. symmetry analysis) do not allow finding exact solutions with prescribed geometrical properties of the governed flow. The goal of this paper is to develop a suitable framework for the description of MHD flows with non-trivial topology, and to construct new classes of exact solutions that possess functional arbitrariness and govern non-stationary plasma flows with knotted configurations of magnetic lines and magnetic surfaces.

%%
%%The most universal method for constructing exact solutions to differential equations is the symmetry analysis  \cite{LVO1982en, Olver2000en}. Group of symmetries for MHD equations was first calculated in \cite{Fuchs1987}. A number of paper devoted to construction of exact invariant and conditionally invariant symmetry reductions were published since that time \cite{DonatoOliveri1993, Galas1993, Fuchs1991, GalasRichter1991, GrundLalag1995, TTH2002, TTH2003, Picard2008, GrundPicard2004}. Exact solutions of the Beltrami type were extensively studied in  \cite{Bogoyavl2000, Bogoyavl2000Comptes, Bogoyavl2003, BogoyavlFuchssteiner2004}. These solutions were generalized to MHD equations with anisotropic stress tensor in \cite{BogoyavlChev2004, Chevyakov2005}. Solutions of ideal MHD equations with homogenious deformation (velocity linearly depend on spatial coordinates) were investigated in \cite{Kulik1958, Naumov2001}. Papers \cite{Golovin2005-1, Golovin2005-2, Golovin2006, Golovin2008-1, Golovin2008-2} are devoted to the systematic construction and description of partially invariant solution to MHD equation. The classification of solutions was based on the notion of the hierarchy of partially invariant solutions introduced in \cite{Golovin2008-5}. Unfortunately, due to the method of construction based on invariants of the admissible group most of these solutions can not be restricted to a finite area in 3D space unless some artificial sinks or sources of fluid over the particular surfaces limiting the area of the flow were introduced.

The main idea of the paper is to ``hide'' the geometry of the flow into the special choice of a curvilinear system of coordinates. Namely, the system of coordinates is chosen such that trajectories and magnetic lines of the flow form two families of coordinate curves. This approach allows one to separate the topological structure of the magnetic field and the evolution of the flow. The obtained system of coordinates is referred to as ``natural'' by the analogy with the natural coordinates in classical mechanics. In the natural coordinates the system of MHD equations is reduced to a non-linear vector wave equation and to the incompressibility condition in the form of generalized Cauchy integral. The symmetry group of the obtained system of equations, equivalence transformations, and a group classification with respect to the constitutive equation of fluid are calculated.

The obtained system of equations allows construction of a class of exact solutions describing non-stationary incompressible MHD flows with constant total pressure. The complete description of this class of solutions requires separation of variables in a certain scalar equation with the subsequent solution of an overdetermined system of partially differential equations. This procedure in particular cases gives new exact solutions of MHD equations possessing a functional arbitrariness up to one function of three variables, two functions of two variables and one function of one variable. The solutions describe non-stationary plasma flows with cylindrical, toroidal, or knotted magnetic surfaces.

The stationary case was investigated in the similar manner in \cite{Golovin2010}. Remarkably, that in the analogous natural system of coordinates where streamlines and magnetic lines form two families of coordinate curves, MHD equations were also reduced to a vector wave equation subjected to a geometrical constraint. The class of solutions with constant total pressure was also described. It was shown, that ``Maxwellian'' surfaces that bear magnetic lines and streamlines of the flow belong to the class of translational surfaces obtained by sliding one 3D curve along another. Similar system of coordinates for stationary MHD equations was used by C. Rogers and W. K. Schief \cite{RogersSchief2003} where the relation between solutions in which the total pressure is constant on Maxwellian surfaces with the integrable Pohlmeyer--Lund--Regge model was discovered. The same method was applied to the solution of Gilbarg problem \cite{Gilbarg1947} analogue: to which extent the geometry of the flow (streamlines and magnetic lines pattern) determines the flow, and for the construction of one-parameter family of exact solutions describing an equilibria state with toric geometry \cite{Schief2003}. In the non-stationary case the similar system of curvilinear coordinates constructed on trajectories and magnetic lines of the flow was used by J. D. Gibbon and D. D. Holm  \cite{GibbonHolm2007} for the analysis of alignment dynamics of non-stationary MHD flow using a quaternionic approach.

\section{Preliminary calculations} Ideal magneto-gasdynamics equations have the following dimensionless form:
\begin{subequations}\label{MHD}
\begin{eqnarray}\label{MHDcont}
&&\rho_t+\bu\cdot\nabla\rho+\rho\dv \bu=0,\\[2mm]\label{MHDmom}
&&\rho\Bigl(\bu_t+(\bu\cdot\nabla)\bu\Bigr)+\,\bB\times\rot\bB+\nabla p=0,\\[2mm]\label{MHDind}
&&\bB_t=\rot(\bu\times\bB),\\[2mm]\label{MHDFarad}
&&\dv\bB=0.
\end{eqnarray}
\end{subequations}
Here $\bu$ is the velocity vector, $\bB$ is the magnetic field, $\rho$ is the density, and $p$ is the pressure. System of equations \eqref{MHD} in the case of compressible fluid is extended by the state equation $p=F(\rho,S)$ with entropy $S$ that is conserved along particles' trajectories:
\[S_t+(\bu\cdot\nabla)S=0.\]
For the incompressible fluid $\rho=\const$ the state equation is not required. Pressure $p$ acts as an unknown function.

The system of MHD equations \eqref{MHD} is transformed as follows. By using the vector identity
\[\bB\times\rot\bB=\frac{1}{2}\nabla |\bB|^2-(\bB\cdot\nabla)\bB,\]
and introducing the total pressure $P=p+\frac{1}{2}|\bB|^2$ the momentum equation \eqref{MHDmom} is brought to the form
\[\rho\bigl(\bu_t+(\bu\cdot\nabla)\bu\bigr)-(\bB\cdot\nabla)\bB+\nabla P=0.\]
With the use of an identity
\[\rot(\bu\times\bB)=\bu\dv\bB-\bB\dv\bu+(\bB\cdot\nabla)\bu-(\bu\cdot\nabla)\bB,\]
and equations \eqref{MHDcont}, \eqref{MHDFarad} equation \eqref{MHDind} is modified as follows:
\[\bB_t=\frac{\bB}{\rho}\Bigl(\rho_t+\bu\cdot\nabla\rho\Bigr)+(\bB\cdot\nabla)\bu-(\bu\cdot\nabla)\bB.\]
The latter can be conveniently written as
\begin{equation}\label{commut}
\Dt \left(\frac{\bB}{\rho}\right)=\Ds\bu
\end{equation}
where notations
\[\Dt=\partial_t+\bu\cdot\nabla, \quad\Ds=\frac{\bB}{\rho}\cdot\nabla\]
for differentiations along trajectories and magnetic lines respectively are introduced. Note, that equation \eqref{commut} is equivalent to the commutativity of vector fields $\Dt$ and $\Ds$ in $\mathbb{R}^4(t,\bx)$.

Finally, with the use of the modified magnetic field $\bb=\rho^{-1}\bB$ equations \eqref{MHD} are written as follows:
\begin{subequations}\label{MHD1}
\begin{eqnarray}\label{MHD1cont}
&&\rho_t+\bu\cdot\nabla\rho+\rho\dv \bu=0,\\[2mm]\label{MHD1mom}
&&\rho\Bigl(\Dt\bu-\Ds(\rho\bb)\Bigr)+\nabla P=0,\\[2mm]\label{MHD1ind}
&&\Dt\bb=\Ds\bu,\\[2mm]\label{MHD1Farad}
&&\dv(\rho\bb)=0.
\end{eqnarray}
\end{subequations}
This form of MHD equations will be used for construction of the curvilinear coordinate system in the following sections.

\section{Natural curvilinear system of coordinates}\label{sec2} A curvilinear coordinate system $(t,\bxi)$, determined for each fixed $t$ by the diffeomorphism $\bx=\bgamma(t,\bxi)$  is introduced over an open subset of the Euclidean space $\mathbb{R}^4(t,\bx)$. Basic vectors of the coordinate system are defined at each point as
\[\be_0=(1,\bgamma_t),\quad\be_i=(0,\bgamma_{\xi^i}),\quad i=1,2,3.\]
As before, the lower indices denote the corresponding partial derivatives. Commutation relation \eqref{commut} for vector fields $\Dt$ and $\Ds$ allows choosing these fields as basic ones \cite{Schutz1980en}:
\[\be_0=(1,\bu),\quad \be_1=(0,\bb).\]
The two remaining vector fields $\be_2$ and $\be_3$ are not fixed. Thus, coordinate lines of $t$-family coincide with particles' trajectories; Coordinate lines of $\xi^1$-family trace the magnetic lines. At that, relation \eqref{commut} is equivalent to the equality of mixed derivatives of function $\bgamma$:
\[\Ds\bu=\pd{}{\xi^1}\pd{\bgamma}{t}=\pdd{\bgamma}{t}{\xi^1}=\pd{}{t}\pd{\bgamma}{\xi^1}=\Dt\bb.\]
In what follows the notation $\xi^0=t$, and the Einstein convention of summation with respect to the repeating indices are adopted. At that, if the summation is performed with respect to a Greek index, it is supposed to take values from 0 to 3; The repeated Roman index runs from 1 to 3.
Contravariant components of vectors $\bu=u^\alpha\be_\alpha$ and $\bb=b^\alpha\be_\alpha$ in the curvilinear coordinates take the following simple form:
\begin{equation}\label{coords}
u^0=1,\quad u^i=0,\quad i=1,2,3;\quad b^0=0,\quad b^1=1,\quad b^2=b^3=0.
\end{equation}

The constructed curvilinear system of coordinates will be referred to as `natural' by the analogy with natural coordinates in classical mechanics. The natural system of coordinates can also be regarded as the special choice of Lagrangian coordinates. The arbitrariness in the choice of Lagrangian coordinates $\bxi$ is used for ``straightening'' of magnetic field lines. Thus, the complexity of the magnetic field is hidden in the proper choice of the coordinate system. The initial magnetic field specifies the coordinate system in the initial state. Time evolution of the magnetic field automatically follows from the evolution of the coordinate system due to the frozenness of the magnetic field into the flow.

\section{Equations in natural coordinates}\label{sec3} In this section equations \eqref{MHD1} will be written in the natural coordinates with the use of formulae given in Appendix \ref{A1}. By virtue of expression \eqref{coords} one has
\[\dv\bu=u^i_{,\,i}=u^\beta\Gamma_{\beta \,i}^i=\Gamma_{0i}^i=\frac{1}{\sqrt{g}}\pd{\sqrt{g}}{t}.\]
Here $f_{,i}$ denotes the covariant derivative of function f with respect to $i$-th variable; $g$ is the determinant of the metric tensor; $\Gamma_{\alpha\beta}^\kappa$ are Christoffel symbols of second kind. Equation \eqref{MHD1cont} is transformed as follows
\begin{equation}\label{MHD2cont}
\pd{\rho}{t}+\frac{\rho}{\sqrt{g}}\pd{\sqrt{g}}{t}=0\quad\Rightarrow \quad\pd{}{t}\ln\Bigl(\rho\sqrt{g}\Bigr)=0.
\end{equation}
In the same way, equation \eqref{MHD1Farad} can be written as
\begin{multline}\label{MHD2Farad}
(\rho \,b^i)_{,\,i}=\pd{\rho \,b^i}{\xi^i}+\rho \,b^\alpha\Gamma_{\alpha i}^i=\pd{\rho}{\xi^1}+\rho\Gamma_{1i}^i\\[2mm]
=\pd{\rho}{\xi^1}+\frac{\rho}{\sqrt{g}}\pd{\sqrt{g}}{\xi^1}=\rho\left(\pd{}{\xi^1}\ln\Bigl(\rho\sqrt{g}\Bigr)\right)=0.
\end{multline}
Equations \eqref{MHD2cont} and \eqref{MHD2Farad} lead to the generalized Cauchy integral
\begin{equation}\label{Cauchy}
\rho\sqrt{g}=f(\xi^2,\xi^3)
\end{equation}
with an arbitrary function $f$.

It is remained to rewrite the momentum equation \eqref{MHD1mom}. The analogous computation of covariant derivatives taking into account formulae \eqref{coords} gives
\begin{equation}\label{MHD2mom}
\rho\left(\Gamma_{00}^i-\pd{\rho}{\xi^1}-\rho\Gamma_{11}^i\right)+g^{ik}\pd{P}{\xi^k}=0,\quad i=1,2,3.
\end{equation}
Note the following identities:
\[\Gamma_{00}^i\be_i=\pd{\be_0}{t},\quad \Gamma_{11}^i\be_i=\pd{\be_1}{\xi^1},\quad \be_i\,g^{ik} =\be^k\]
Multiplication of equations \eqref{MHD2mom} on basic vectors $\be_i$ and summation with respect to $i$ in view of the preceding identities gives
\[
\rho\left(\ppd{\bgamma}{t}-\pd{}{\xi^1}\left(\rho\pd{\bgamma}{\xi^1}\right)\right)+ \left(\pd{\bxi}{\bgamma}\right)^T\nabla_{\xi}P=0.
\]
Hereafter symbol ``$T$'' denotes the transpose of the matrix; $(\partial\bxi/\partial\bgamma)$ is Jacobian matrix. Finally, equations for the sought mapping $\bx=\bgamma(t,\bxi)$ take the following form:
\begin{equation}\label{main}
\begin{array}{l}
\displaystyle \left(\pd{\bgamma}{\bxi}\right)^T\left(\ppd{\bgamma}{t}-\pd{}{\xi^1}\left(\rho\pd{\bgamma}{\xi^1}\right)\right)+
\frac{1}{\rho}\,\nabla_{\xi}P=0,\\[4mm]
\displaystyle \rho\sqrt{g}=f(\xi^2,\xi^3),\quad \sqrt{g}=\det\left(\pd{\bgamma}{\bxi}\right).
\end{array}
\end{equation}
In the case of the incompressible fluid ($\rho=\rho_0$) function $P$ is treated as an unknown function. For the compressible fluid with the state equation $p=F(\rho,S)$ relations \eqref{main} are extended by the expression for the complete pressure in terms of density and entropy as
\begin{equation}\label{Peqn}
P=F\bigl(\rho,S(\bxi)\bigr)+\frac{1}{2}\,\rho^2\left|\pd{\bgamma}{\xi^1}\right|^2
\end{equation}
with an arbitrary function $S(\bxi)$. Thus, in the natural system of coordinates original MHD equations \eqref{MHD} are reduced to the nonlinear vector wave equation and to the incompressibility condition in the form of the generalized Cauchy integral.

\section{The physical picture of the flow and initial conditions} Each solution $\bx=\bgamma(t,\bxi)$ of the system \eqref{main}, \eqref{Peqn} provides an explicit description of particles' trajectories and magnetic lines of the governed fluid flow.

Particle trajectory originating at time $t=0$ at the position $\bx_0=\bgamma(0,\bxi_0)$ is given explicitly by the parametric formula $\bx(t)=\bgamma(t,\bxi_0)$. Magnetic lines are parameterized by coordinate $\xi^1$. At the arbitrary moment of time $t_0$ the parametrization of a magnetic line has the form $\bx(s)=\bgamma(t_0,s,\xi^2_0,\xi^3_0)$. At that, coordinates $\xi^2_0$ and $\xi^3_0$ ``distinguish'' different magnetic lines.

Note, that the initial function $\bgamma(0,\bxi)=\bgamma_0(\bxi)$ can not be taken arbitrarily. Indeed, let the initial density distribution $\rho(0,\bx)=\rho_0(\bx)$ be given. Then, the dependence $\bgamma_0(\bxi)$ should satisfy the equation
\begin{equation}\label{initcond}
\rho_0\bigl(\bgamma_0(\bxi)\bigr)\det\left(\pd{\bgamma_0}{\bxi}\right)=f(\xi^2,\xi^3)
\end{equation}
which is equivalent to the incompressibility of the initial magnetic field stated by equation \eqref{MHDFarad}.

Suppose the divergent-free magnetic field $\bB_0(\bx)$, density $\rho_0(\bx)$, and initial velocity field $\bu_0(\bx)$ of fluid be given at initial time $t=0$. Hence, the magnetic field uniquely covers the area of the flow such that only one magnetic line passes through each point. Construction of the initial data for system \eqref{main} is performed as follows.

With the vector field $\bb_0(\bx)=\rho_0^{-1}(\bx)\bB_0(\bx)$ one solves the system of ordinary differential equations
\[
\opd{\bx}{\xi^1}=\bb_0(\bx).
\]
The general solution of this system contains three arbitrary constants. One of these constants corresponds to the trivial admissible translational symmetry $\xi^1\rightarrow \xi^1+a$ and is not important. The two remaining constants are denoted as $\xi^2$ and $\xi^3$. As a result, this gives a diffeomorphism $\bx=\bgamma_0(\bxi)$ (at least, for smooth vector field $\bb_0$). Moreover, due to the divergence-freeness of the initial vector field $\bB_0(\bx)$ this diffeomorphism automatically satisfies the incompressibility condition \eqref{initcond}.

For the given initial velocity field $\bu_0(\bx)$, the vector field
\[\bgamma_1(\bxi)=\bu_0\bigl(\bgamma_0(\bxi)\bigr)\]
is constructed. This finally gives the initial conditions for the system of equations \eqref{main} in the form
\[\bgamma(0,\bxi)=\bgamma_0(\bxi),\quad \pd{\bgamma}{t}(0,\bxi)=\bgamma_1(\bxi).\]
In case of incompressible fluid these data are extended by specification of the initial pressure:
\[P(0,\bxi)=p_0\bigl(\gamma_0(\bxi)\bigr)+\frac{1}{2}\,\Bigl|B_0\bigl(\gamma_0(\bxi)\bigr)\Bigr|^2.\]

\section{Symmetries and equivalence transformations} System of equations \eqref{main} is related to the initial MHD system in Cartesian coordinates by the non-local transformation and partial integration. Hence, the admissible group of symmetries is not necessarily conserves. In this section symmetry properties of equation \eqref{main} are investigated.

Equations \eqref{main} admit the following equivalence transformation
\begin{equation}\label{Eqtrans}
\overline{\xi}^2=\overline{\xi}^2(\xi^2,\xi^3),\quad \overline{\xi}^3=\overline{\xi}^3(\xi^2,\xi^3), \quad \overline{f}=f/\Delta,\quad \Delta=\pd{(\overline{\xi}^2,\overline{\xi}^3)}{(\xi^2,\xi^3)}\ne0.
\end{equation}
Here $\partial(\cdot,\cdot)/\partial(\cdot,\cdot)$ denotes the corresponding Jacobian. Since $f\ne0$, by the suitable choice of functions $\overline{\xi}^2(\xi^2,\xi^3)$ and $\overline{\xi}^3(\xi^2,\xi^3)$ one can make $f=1$.
In addition to the equivalence transformation \eqref{Eqtrans}, system \eqref{main} always admits the ``reparametrization'' symmetry
\begin{equation}\label{ReparamSymm}
\overline{\xi}^1=\xi^1+a(\xi^2,\xi^3),\quad \overline{\xi}^2=\overline{\xi}^2(\xi^2,\xi^3),\quad \overline{\xi}^3=\overline{\xi}^3(\xi^2,\xi^3),\quad \pd{(\overline{\xi}^2,\overline{\xi}^3)}{(\xi^2,\xi^3)}=1.
\end{equation}
with arbitrary smooth functions $a$, $\overline{\xi}^2$ and $\overline{\xi}^3$.

Symmetries of system \eqref{main} are calculated separately for cases of compressible and incompressible fluids. In the incompressible case the density is normalized to unity: $\rho=1$. In order to have an explicit expression of the time derivative it is convenient to rewrite system \eqref{main} in the following form:
\begin{equation}\label{MainIncompr}
\begin{array}{l}
\displaystyle \ppd{\bgamma}{t}-\ppd{\bgamma}{\xi^1}+
\pd{P}{\xi^1}\left(\pd{\bgamma}{\xi^2}\times\pd{\bgamma}{\xi^3}\right) +
\pd{P}{\xi^2}\left(\pd{\bgamma}{\xi^3}\times\pd{\bgamma}{\xi^1}\right) \\[4mm]
\displaystyle \phantom{\ppd{\bgamma}{t}-}+\pd{P}{\xi^3}\left(\pd{\bgamma}{\xi^1}\times\pd{\bgamma}{\xi^2}\right)=0,\\[4mm]
\displaystyle \left[\pd{\bgamma}{\xi^1},\pd{\bgamma}{\xi^2},\pd{\bgamma}{\xi^3}\right]=1.
\end{array}
\end{equation}
Here $[\cdot,\cdot,\cdot]$ denotes the mixed product of three vectors; $\times$ is the vector product in $\mathbb{R}^3$.

The direct computation of the symmetry group of system \eqref{MainIncompr} using the standard algorithm \cite{LVO1982en,Olver2000en} gives the following result.

\begin{theorem}\label{th1} The symmetry group of system \eqref{MainIncompr} is a semidirect product of the reparametrization group \eqref{ReparamSymm} and the generalized Galilean group. The latter is generated by the following infinitesimal operators:
\[
\begin{array}{l}
T=\partial_t,\\[2mm]
R_{i}=\varepsilon_{ijk}\gamma^j\partial_{\gamma^k},\quad i=1,2,3, \\[2mm]
S_1=t\partial_t+\xi^j\partial_{\xi^j}+\gamma^j\partial_{\gamma^j},\\[2mm]
S_2=3\xi^2\partial_{\xi^2}+3\xi^3\partial_{\xi^3}+2\gamma^j\partial_{\gamma^j}+4P\partial_P,\\[2mm]
G_i(\alpha^i(t))=\alpha^i(t)\partial_{\gamma^i}-\gamma^i{\ddot{\alpha}^i}(t)\partial_P,\, i=1,2,3,\\[2mm]
K=\beta(t)\partial_P.
\end{array}
\]
Here the summation is taken only with respect to $j,k=1,2,3$; $\alpha^i(t)$ and $\beta(t)$ are arbitrary smooth functions; $\epsilon_{ijk}$ is the Levi-Civita symbol; upper dot denotes the $t$-derivative.
\end{theorem}

Note that the admissible group, except for its ``reparametrization'' part \eqref{ReparamSymm}, is inherited from the original MHD equations \eqref{MHD} in Eulerian coordinates. Operator $T$ generates a time shift, operators $R_i$ give the group of rotations in $\mathbb{R}^3(\bgamma)$, $S_1$ and $S_2$ generate dilatations, operator $K$ claims that pressure can be modified by addition of an arbitrary function of time. The most non-trivial transformation is the generalized Galilean translation generated by operators $G_i(\alpha^i(t))$
\[\overline{\bgamma}=\bgamma+\balpha(t),\quad \overline{P}=P-\bgamma\cdot\ddot{\balpha}(t)- \frac{1}{2}\,\balpha(t)\cdot\ddot{\balpha}(t)\]
with an arbitrary time-dependent vector-function $\balpha(t)$.

In the case of compressible fluid with the constitutive equation $p=F(\rho,S)$ relations \eqref{main} can be conveniently written as follows:
\begin{equation}\label{MainCompr}
\begin{array}{l}
\displaystyle \ppd{\bgamma}{t}-\pd{}{\xi^1}\left(\rho\pd{\bgamma}{\xi^1}\right)+
\pd{P}{\xi^1}\left(\pd{\bgamma}{\xi^2}\times\pd{\bgamma}{\xi^3}\right) +
\pd{P}{\xi^2}\left(\pd{\bgamma}{\xi^3}\times\pd{\bgamma}{\xi^1}\right) \\[4mm]
\displaystyle \phantom{\ppd{\bgamma}{t}-}+\pd{P}{\xi^3}\left(\pd{\bgamma}{\xi^1}\times\pd{\bgamma}{\xi^2}\right)=0,\\[4mm]
\displaystyle \rho\left[\pd{\bgamma}{\xi^1},\pd{\bgamma}{\xi^2},\pd{\bgamma}{\xi^3}\right]=1,\quad
\displaystyle \pd{p}{t}=h(p,\rho)\pd{\rho}{t},\quad P=p+\frac{1}{2}\,\rho^2\left|\pd{\bgamma}{\xi^1}\right|^2.
\end{array}
\end{equation}
Here $h(p,\rho)$ is a square of the speed of sound $h(p,\rho)=\pd{F}{\rho}|_{S=S(p,\rho)}$. Unknown functions in system \eqref{MainCompr} are $\bgamma$, $p$ and $\rho$.

Computation of the symmetry group of system \eqref{MainCompr} requires the group classification with respect to the constitutive equation represented by function $h(p,\rho)$. The system admits the following equivalence transformations:
\begin{equation}\label{EquivTrans}
\begin{array}{c}
\overline{t}=\alpha^3 t,\quad \overline{\xi}^1=\xi^1,\quad \overline{\xi}^2=\beta^3\xi^2, \quad\overline{\xi}^3=\beta^3\xi^3,\quad \overline{\bgamma}=\beta^2\bgamma,\quad \overline{\rho}=\alpha^{-6}\rho,\\[2mm]
\overline{p}=\alpha^{-8} \beta^4p+\kappa,\quad\overline{h}\bigl(\overline{p},\overline{\rho}\bigr)=\alpha^{-2}\beta^4 h\bigl(\alpha^8\beta^{-4}(\overline{p}-\kappa),\alpha^6\overline{\rho}\bigr)
\end{array}
\end{equation}
with arbitrary constants $\alpha$, $\beta$, and $\kappa$. The result of calculation of the symmetry group is the following.

\begin{theorem}\label{th2}
The kernel of admissible group of equations \eqref{MainCompr} is a semidirect sum of the infinite-dimensional reparametrization group \eqref{ReparamSymm} and the extended Galilean group generated by the following infinitesimal operators:
\[
\begin{array}{l}
T_0=\partial_t,\\[2mm]
T_i=\partial_{\gamma^i},\quad i=1,2,3,\\[2mm]
R_{i}=\varepsilon_{ijk}\gamma^j\partial_{\gamma^k},\quad i=1,2,3,\\[2mm]
S=t\partial_t+\xi^j\partial_{\xi^j}+\gamma^j\partial_{\gamma^j},\\[2mm]
G_i=t\partial_{\gamma^i},\quad i=1,2,3.
\end{array}
\]
As before, the summation is taken only with respect to $j,k=1,2,3$; $\epsilon_{ijk}$ is the Levi-Civita symbol. This group is admitted for an arbitrary function $h(p,\rho)$. For special functions $h$ the kernel of admissible group is extended by the family of operators
\begin{multline*}
Y=c_1 t\partial_t+(-4c_1+2c_2)\xi^1\partial_{\xi^1}+c_2\xi^2\partial_{\xi^2}+c_2\xi^3\partial_{\xi^3} \\[2mm] +2c_1\gamma^j\partial_{\gamma^j}+\bigl(c_3+(4c_2-8c_1)p\bigr)\partial_p+(4c_2-10c_1)\rho\partial_\rho
\end{multline*}
provided the following equation is identically satisfied for $p$ and $\rho$
\begin{equation}\label{ClassEq}
2c_1(h+4ph_p+5\rho h_\rho)-4c_2(ph_p+\rho h_\rho)-c_3 h_p=0.
\end{equation}
Here $c_i$ are arbitrary constants.
\end{theorem}

\begin{table}
\centering
\caption{The result of group classification}\label{t1}
\begin{tabular}{c|c|c|c}
  \hline
    & $h(p,\rho)$ & $F(\rho,S)$ & Group extension \\\hline
  1 & 0 & $F(S)$ & $Y_1$, $Y_2$, $Y_3$ \\[2mm]
  2 & $\rho^k$, $k\ne -1$ & $\frac{1}{k+1}\,\rho^{k+1}+F(S)$ & $2k Y_1+(1+5k)Y_2,\, Y_3$ \\[2mm]
  3 & $\rho^{-1}$ & $\ln\rho+F(S)$ & $Y_1 +2Y_2,\, Y_3$ \\[2mm]
  4 & $k p/\rho$, $k\ne0$ & $F(S)\rho^k$ & $Y_1$, $Y_2$ \\[2mm]
  5 & $p^k f\bigl(\rho\,p^{k-1}\bigr)$ & $g\bigl(\rho\,p^{k-1}\bigr)F(S)$ & $2k Y_1+(1+4k)Y_2$ \\[2mm]
  6 & $f\bigl(\rho e^p\bigr)/\rho$ & $g\bigl(\rho\,e^p\bigr)+F(S) $ & $Y_1+2Y_2+2Y_3$ \\[2mm]
  7 & $f(\rho)$ & $g(\rho)+F(S)$ & $Y_3$\\[2mm]
  8 & $f(p)/\rho$ & $g\bigl(\rho F(S)\bigr) $ & $Y_1+2Y_2$\\[2mm]
  9 & $h(p,\rho)$ & $F(\rho,S)$ & ---
\end{tabular}
\end{table}

Group classification with respect to $h(p,\rho)$ is performed in the same way as in classical gas dynamics \cite{LVO1982en}. The result is summarized in Table \ref{t1}. Here $f$, $F$, and $g$ are arbitrary functions, $k$ is an arbitrary real parameter. The first column represents the order number of the group extension. The second column gives the particular form of function $h(p,\rho)$ normalized by the equivalence transformations \eqref{EquivTrans}. In the third column the corresponding constitutive equation $p=F(\rho,S)$ is specified. The last column of the Table provides infinitesimal operators of transformations, that extend the kernel of the admissible group for the particular constitutive equation. The following notations for the extension operators are used:
\[
\begin{array}{l}
Y_1=t\partial_t-4\xi^1\partial_{\xi^1}+2\gamma^j\partial_{\gamma^j}-8p\partial_p-10\rho\partial_\rho,\\[2mm]
Y_2=2\xi^1\partial_{\xi^1}+\xi^2\partial_{\xi^2}+\xi^3\partial_{\xi^3}+4p\partial_p+4\rho\partial_\rho,\quad
Y_3=\partial_p.
\end{array}
\]
Operators $Y_1$ and $Y_2$ correspond to dilatations, operator $Y_3$ generates the shift of pressure. Details of the group classification can be found in Appendix \ref{A2}.

\section{Flows with constant total pressure} This section is devoted to the construction of a special class of solutions to equations \eqref{MainIncompr} of incompressible infinitely conducting fluid. Suppose that the total pressure is constant over the whole area occupied by the flow:
\[P=p+\frac{1}{2}\,\left|\bB\right|^2=\const.\]
From the first of equations \eqref{MainIncompr} it follows that
\[\ppd{\bgamma}{t}-\ppd{\bgamma}{\xi^1}=0.\]
Integration of this equation yields the expression for function $\bgamma(t,\bxi)$:
\begin{equation}\label{gammarepr}
\bgamma=\bsigma(t-\xi^1,\xi^2,\xi^3)+\btau(t+\xi^1,\xi^2,\xi^3).
\end{equation}
With this representation the incompressibility condition (the second of equations \eqref{MainIncompr}) must be satisfied:
\begin{equation}\label{varsep}
(-\bsigma_1+\btau_1)\cdot\Bigl((\bsigma_2+\btau_2)\times(\bsigma_3+\btau_3)\Bigr)=1.
\end{equation}
Hereafter the lower index `$i$' denotes the derivative with respect to $i$-th argument of functions $\bsigma$ and $\btau$. In the expanded form scalar equation \eqref{varsep} contains 48 additives. Each of them has the form of a product of two multipliers depending on different variables: $t-\xi^1$ and $t+\xi^1$. Thus, it is required to separate variables in equation \eqref{varsep} and then to solve the obtained overdetermined systems of nonlinear partially differential equations for functions $\btau$ and $\bsigma$. Not pretending to solve this complicated problem in the general form, two nontrivial examples of exact solutions are constructed below.

\section{Stationary field-aligned flows} Suppose that $\bsigma\equiv0$ over the whole area of the flow. The solution is any vector field $\btau$ with a unit Jacobian:
\begin{equation}\label{fieldalligned}
\bgamma(t,\bxi)=\btau(t+\xi^1,\xi^2,\xi^3),\quad \det\left(\pd{\btau}{\bxi}\right)=1.
\end{equation}
In this solution velocity vector $\bu$ and magnetic field $\bB$ are collinear. The solution describes stationary flow  of plasma aligned with the magnetic field known as Chandrasekhar's solutions \cite{Chandra1956}.

\section{Non-stationary flows}
In this special case it is supposed that vector field $\bsigma$ depends only on one variable $t-\xi^1$:
\begin{equation}\label{specrepr}
\bgamma=\bsigma(t-\xi^1)+\btau(t+\xi^1,\xi^2,\xi^3).
\end{equation}
Equation (\ref{varsep}) is reduced to the following one:
\begin{equation}\label{varsepspec}
-\bsigma'\cdot(\btau_2\times\btau_3)=1-\btau_1\cdot(\btau_2\times\btau_3).
\end{equation}
Prime denotes the derivative of vector field $\bsigma$ with respect to its only argument.

Henceforth notations $\bbeta^1$, $\bbeta^2$, and $\bbeta^3$ for vectors of standard Cartesian basis in  $\mathbb{R}^3$ are adopted. Vector field $\btau$ has the following decomposition: $\btau=\sum_{i=1}^3\tau^i\bbeta^i$. The solution of equation (\ref{varsepspec}) depends on the dimension $\dim\{\bsigma\}$ of the linear space spanned by the vector $\bsigma'(t-\xi^1)$ as its argument varies.

\underline{$\dim\{\bsigma\}=1$}. Without loss of generality it is supposed that vector $\bsigma'$ is equal to $\bbeta^1$. Separation of variables in equation (\ref{varsepspec}) gives
\[\bsigma'=\bbeta^1,\quad (\btau_1-\bbeta^1)\cdot(\btau_2\times\btau_3)=1.\]
Integration of the first equation with respect to $t-\xi^1$ accurate to the insufficient constant additive vector yields
\[\bsigma=(t-\xi^1)\bbeta^1.\]
It is convenient to represent vector $\btau$ as $\btau=(t+\xi^1)\bbeta^1+\widetilde{\btau}$. Then the solution is reduced to (tilde is dropped)
\[\bgamma=2t\,\bbeta^1+\btau(t+\xi^1,\xi^2,\xi^3),\quad \det\left(\pd{\btau}{\bxi}\right)=1.\]
Now we see that accurate to the Galilean translation along $\bbeta^1$ this solution coincides with the stationary solution given by (\ref{fieldalligned}).

\underline{$\dim\{\bsigma\}=2$}. Separation of variables in equation (\ref{varsepspec}) gives
\begin{multline}\label{s1}
\bsigma'=U(t-\xi^1)\bbeta^1+\bbeta^2,\quad \bbeta^1\cdot(\btau_2\times\btau_3)=0,\\ (\btau_1-\bbeta^2)\cdot(\btau_2\times\btau_3)=1.
\end{multline}
Here $U$ is a non-constant function. Integration of the first equation of (\ref{s1}) accurate to the insufficient additive constant vector leads to
\[\bsigma=u(t-\xi^1)\bbeta^1+(t-\xi^1)\bbeta^2.\]
Here $u'(\alpha)=U(\alpha)$. The second of equations (\ref{s1}) implies that components $\tau^2$ and $\tau^3$ of vector $\btau$ are functionally dependent as functions of $\xi^2$ and $\xi^3$:
\[
\tau^2=\tau^2\bigl(t+\xi^1,\lambda(t+\xi^1,\xi^2,\xi^3)\bigr),\quad \tau^3=\tau^3\bigl(t+\xi^1,\lambda(t+\xi^1,\xi^2,\xi^3)\bigr).
\]
It is convenient to change vector $\btau$ as $\btau-(t+\xi^1)\bbeta^2\rightarrow \btau$. At that, vector $\bgamma$ acquires an additive $2t\bbeta^2$, which can be zeroed by the Galilean translation. Finally, vector $\bgamma$ is obtained in the following form:
\begin{multline}\label{sol1}
\bgamma=\Bigl(u(t-\xi^1)+\tau^1(t+\xi^1,\xi^2,\xi^3)\Bigr)\bbeta^1%\\[2mm]
+\tau^2\bigl(t+\xi^1,\lambda(t+\xi^1,\xi^2,\xi^3)\bigr)\bbeta^2\\[2mm]
+\tau^3\bigl(t+\xi^1,\lambda(t+\xi^1,\xi^2,\xi^3)\bigr)\bbeta^3,
\end{multline}
\begin{equation}\label{soleq1}
\left|\begin{array}{ll}
\tau^2_1&\tau^2_\lambda\\[2mm]
\tau^3_1&\tau^3_\lambda
\end{array}\right|
\left|\begin{array}{ll}
\lambda_2&\lambda_3\\[2mm]
\tau^1_2&\tau^1_3
\end{array}\right|=1.
\end{equation}
Scalar equation (\ref{soleq1}) states that the product of two Jacobians must be equal to 1. Note, that without normalization (\ref{Eqtrans}) the right-hand side of this equation is an arbitrary function $f(\xi^2,\xi^3)$. Solution (\ref{sol1}) contains one arbitrary function of three arguments (either of functions $\tau^1$ or $\lambda$), two arbitrary functions of two arguments ($\tau^2$ and $\tau^3$), and one arbitrary function of one argument (function $u$). Equation (\ref{soleq1}) can be treated as a linear equation for $\tau^1$  provided all remaining arbitrary functions are specified.

\underline{$\dim\{\bsigma\}=3$}. Separation of variables in equation (\ref{varsepspec}) gives
\begin{multline}\label{s2}
\bsigma'=U^1(t-\xi^1)\bbeta^1+U^2(t-\xi^1)\bbeta^2+\bbeta^3,\quad \bbeta^1\cdot(\btau_2\times\btau_3)=0,\\[2mm] \bbeta^2\cdot(\btau_2\times\btau_3)=0,\quad (\btau_1-\bbeta_1)\cdot(\btau_2\times\btau_3)=1.
\end{multline}
Here functions $U^1$, $U^2$, and $1$ are supposed to be linearly independent as functions of $t-\xi^1$. Integration of the first of equation (\ref{s2}) gives the expression for vector $\bsigma$:
\[\bsigma=u^1(t-\xi^1)\bbeta^1+u^2(t-\xi^1)\bbeta^2+(t-\xi^1)\bbeta^3.\]
The second and the third equations of (\ref{s2}) are compatible with the last equation in (\ref{s2}) only if $\tau^3$ does not depend on $\xi^2$ and $\xi^3$: $\tau^3=\tau^3(t+\xi^1)$. As before, accurate to the Galilean translation, one obtains the solution:
\begin{multline}\label{sol2}
\bgamma=\Bigl(u^1(t-\xi^1)+\tau^1(t+\xi^1,\xi^2,\xi^3)\Bigr)\bbeta^1\\[2mm]
+\Bigl(u^2(t-\xi^1)+\tau^2(t+\xi^1,\xi^2,\xi^3)\Bigr)\bbeta^2+\tau^3(t+\xi^1)\bbeta^3,
\end{multline}
\begin{equation}\label{soleq2}
{\tau^3}'\left|\begin{array}{ll}
\tau^1_2&\tau^1_3\\[2mm]
\tau^2_2&\tau^2_3
\end{array}
\right|=1.
\end{equation}
This solution contains one arbitrary function of three arguments (either of $\tau^1$ and $\tau^2$), and three arbitrary functions of one argument (functions $\tau^3$, $u^1$ and $u^2$). Scalar equation (\ref{soleq2}) can be treated as a linear equation for $\tau^1$ that also contains an arbitrariness in one function of two arguments. Both solutions can be extended by the application of the admissible symmetry transformations specified in Theorem \ref{th1}.

Solutions (\ref{sol1}) and (\ref{sol2}) can be treated as non-stationary disturbances of the field-aligned flow (\ref{fieldalligned}). The overall geometry of the flow is governed by the vector field $\btau$, while functions $u$ and $u^i$ can be specified arbitrarily and independently providing perturbations of the stationary flow.

\section{Some particular examples of solutions} In this section examples of plasma flows governed by solutions (\ref{sol1}) and (\ref{sol2}) for particular choices of arbitrary functions are constructed.

The first solution under consideration is (\ref{sol2}). For simplicity the normalization  (\ref{Eqtrans}) is not used, so that equation (\ref{soleq2}) reads
\begin{equation}\label{eq2}
\pd{\tau^1}{\xi^2}\pd{\tau^2}{\xi^3}-\pd{\tau^1}{\xi^3}\pd{\tau^2}{\xi^2}=f(\xi^2,\xi^3)
\end{equation}
with an arbitrary function $f$.

On the first step the basic field-aligned stationary solution is constructed by setting $u^1=0$, $u^2=0$. Functions $\tau^1$ and $\tau^3$ are taken as
\begin{equation}\label{sol21}
\begin{array}{l}
\tau^1=a(\mu)+\alpha(\mu)\bigl(A(\xi^2,\xi^3)\cos\varphi(\mu)+B(\xi^2,\xi^3)\sin\varphi(\mu)\bigr),\\[2mm]
\tau^2=b(\mu)-\beta(\mu)\bigl(A(\xi^2,\xi^3)\sin\varphi(\mu)-B(\xi^2,\xi^3)\cos\varphi(\mu)\bigr),\\[2mm]
\mu=t+\xi^1.
\end{array}
\end{equation}
Equation (\ref{eq2}) after separation of variables yields
\begin{equation}\label{eq21}
\opd{\tau^3}{\mu}\,\alpha(\mu)\beta(\mu)=1,\quad f(\xi^2,\xi^3)=\pd{(A,B)}{(\xi^2,\xi^3)}.
\end{equation}
The first equation in (\ref{eq21}) determines functions $\tau^3$ in terms of functions $\alpha$ and $\beta$; The second equation specifies functions $f(\xi^2,\xi^3)$.

Solution (\ref{sol21}) can be treated as follows. Functions $A$ and $B$ can be chosen arbitrarily to select a family of curves in the plane spanned by vectors $\bbeta^1$, $\bbeta^2$. For example, the choice
\[A=k_1 \xi^3\cos\xi^2,\quad B=k_2\xi^3\sin\xi^2\]
specifies a family of nested ellipses parameterized by $\xi^2$ with semi-axes $k_1\xi^3$ and $k_2\xi^3$. This planar picture of curves propagates along $\bbeta^3$ direction being rotated on angle $\varphi(\mu)$, stretched and shifted in $\bbeta^1$ and $\bbeta^2$ directions as specified by functions $\alpha(\mu)$, $\beta(\mu)$, $a(\mu)$, and $b(\mu)$. Since $\alpha$ and $\beta$ are non-zero functions, dependence $\tau^3(\mu)$ is monotonic, therefore variable $\mu$ (or $\xi^1$) can be treated as a parameter along $\bbeta^3$. Magnetic surfaces are obtained by fixing either $\xi^2$ or $\xi^3$. In general, this procedure gives a set of nested deformed cylindrical surfaces for the magnetic surfaces in the basic stationary state.

On the second step one chooses non-zero functions $u^1(t-\xi^1)$, $u^2(t-\xi^1)$ to obtain the non-stationary disturbance of the constructed basic stationary state. As coordinate $\xi^1$ plays a role of a parameter along $\bbeta^3$, the disturbance propagates along $\bbeta^3$ as time grows.

Summing up, solution (\ref{sol2}) describes non-stationary plasma jet flows stretched along one spatial direction.

Now solution (\ref{sol1}) is observed. Again, the normalization (\ref{Eqtrans}) is not performed, so that equation (\ref{soleq1}) reads
\begin{equation}\label{eq12}
\left(\pd{\tau^2}{\xi^1}\pd{\tau^3}{\lambda}-\pd{\tau^2}{\lambda}\pd{\tau^2}{\xi^1}\right)
\left(\pd{\lambda}{\xi^2}\pd{\tau^1}{\xi^3}-\pd{\lambda}{\xi^3}\pd{\tau^1}{\xi^2}\right)=f(\xi^2,\xi^3).
\end{equation}
On the first step the basic stationary solution is constructed by choosing $u=0$. Arbitrary functions in (\ref{sol1}) are selected as follows:
\begin{equation}\label{sol12}
\begin{array}{l}
\tau^1=a(\mu)+B(\xi^2,\xi^3)  \sin\bigl(\varphi(\mu)+A(\xi^2 ,\xi^3)\bigr),\\[3mm]
\lambda =\sqrt{b(\mu)+B(\xi^2,\xi^3)  \cos\bigl(\varphi(\mu)+A(\xi^2 ,\xi^3)\bigr)}\\[3mm]
\tau^2=\lambda\cos k\mu,\quad\tau^3=\lambda\sin k\mu,\quad\mu=t+\xi^1.
\end{array}
\end{equation}
Substitution of (\ref{sol12}) into (\ref{eq12}) gives
\[f(\xi^2,\xi^3)=\frac{k B}{2}\pd{(A,B)}{(\xi^2,\xi^3)},\]
which can be treated as the definition of function $f$. Solution (\ref{sol12}) describes stationary flow with nested toroidal (or knotted) magnetic surfaces provided the inequality $|B(\xi^2,\xi^3)|\le b(\mu)$ is satisfied. At that, variable $\mu$ grows along the central curve of the torus; $\varphi(\mu)$ is responsible for the winding of the magnetic lines around the torus, functions $A$ and $B$ determine the shape of the section of the toroidal surface, functions $a$ and $b$ describe the shift of the section of the torus relative to its central curve.

\begin{figure}
  \centering
  \includegraphics[width=0.5\textwidth]{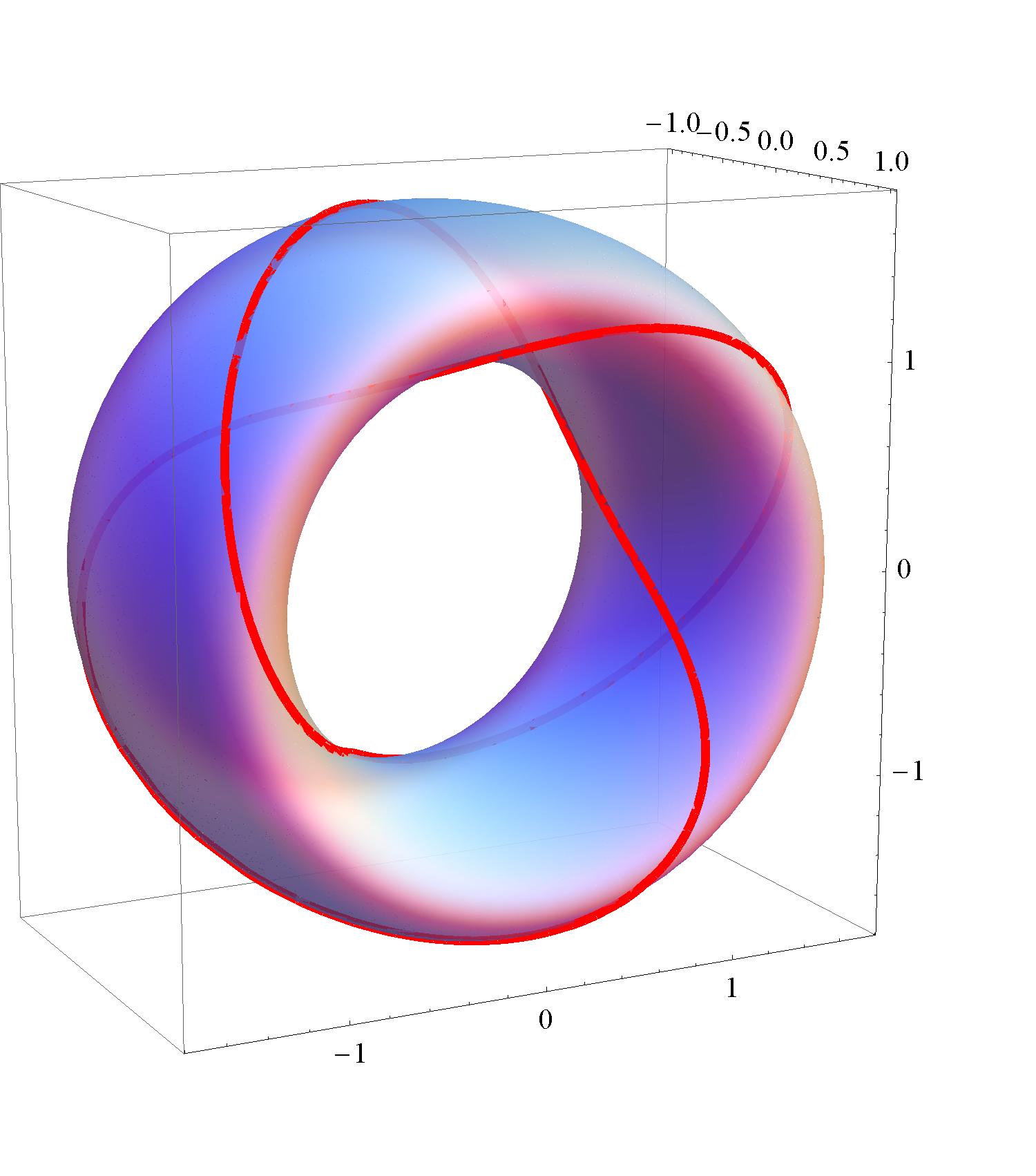}
  \caption{The magnetic surface $\xi^3=1$ with the magnetic line in the shape of the trefoil knot described by solutions (\ref{sol12}), (\ref{sol13}).}\label{F1}
\end{figure}

\begin{figure}
  \centering
  \includegraphics[width=0.5\textwidth]{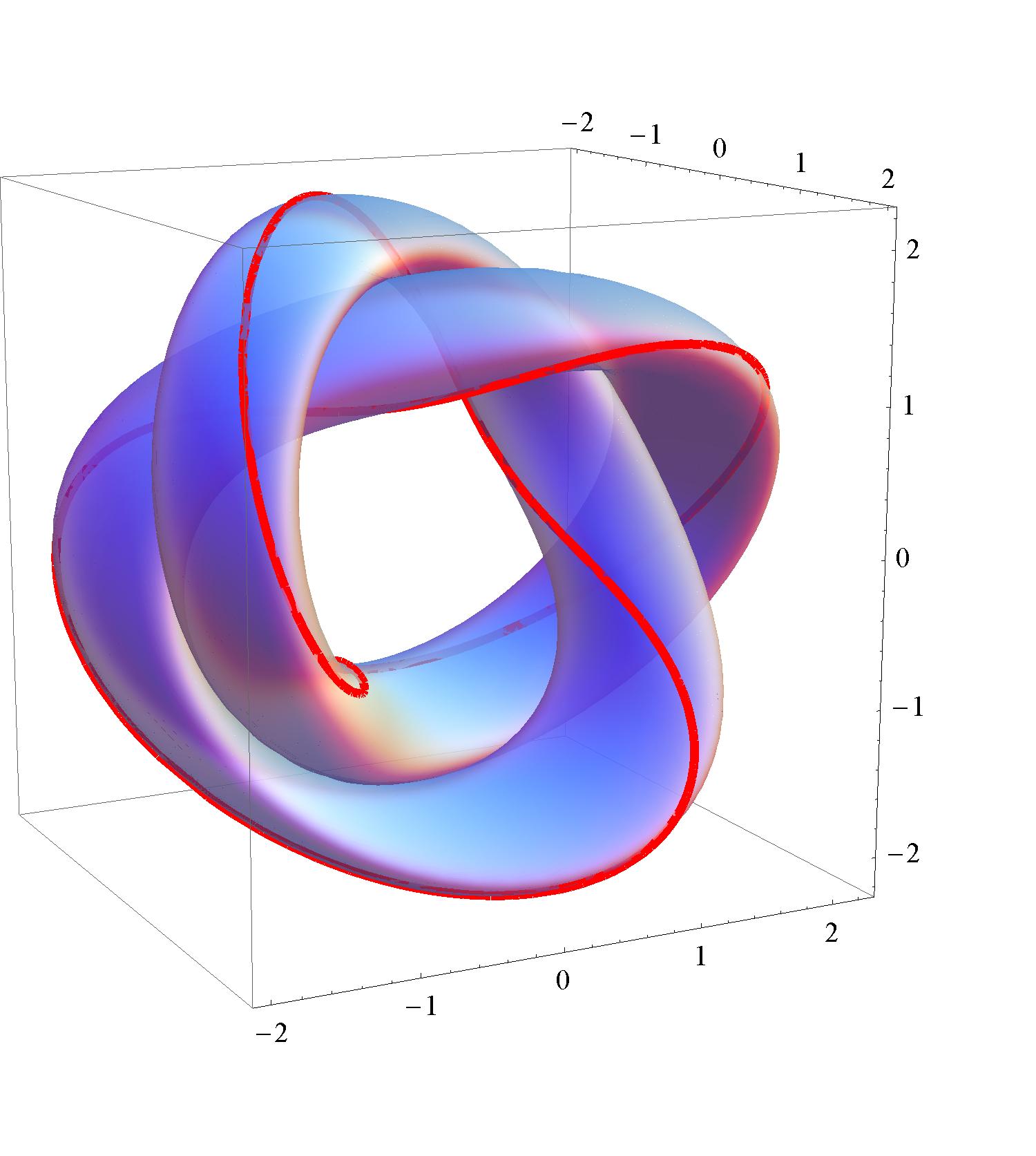}
  \caption{The magnetic surface $\xi^3=1$ in the shape of the trefoil knot described by solutions (\ref{sol12}), (\ref{sol14}).}\label{F2z}
\end{figure}

Suitable choice of arbitrary functions in solution (\ref{sol12}) yields various pictures of magnetic surfaces. For example, the choice
\begin{equation}\label{sol13}
A=\xi^2,\quad B=\xi^3,\quad\varphi=3\mu, \quad a=0,\quad b=2,\quad k=2
\end{equation}
generates the toroidal magnetic surface in Figure \ref{F1}a. All magnetic lines on this surface are closed curves topologically equivalent to the trefoil knot.

The choice
\begin{equation}\label{sol14}
\begin{array}{l}
A=\xi^2,\quad B=\xi^3,\quad\varphi=3\mu, \quad a=\sin3\mu,\\[2mm]
b=3+\cos3\mu,\quad k=2
\end{array}
\end{equation}
leads to the magnetic surfaces in the shape of the trefoil knot depicted in figure \ref{F1}b. One can easily generalize these examples to more topologically complicated pictures of plasma flows.

On the second step one can add a non-stationary disturbance of constructed stationary pictures by choosing a non-zero function $u$. The disturbance propagate along the central curve of the magnetic tubes as time grows. Periodic functions $u$ generate time-periodic disturbances.

As a conclusion, solution (\ref{sol1}) describes various plasma flows with torus- or knot-shaped magnetic surfaces.

\section{Acknowledgements} The work was partially supported by RFBR (grant no. 08-01-00047), by the Ministry of Education and Science of Russian Federation (project no. 02.740.11.0617), and by the Russian Academy of Sciences (project 14.14.1).

\appendix
\section{Some useful formulae}\label{A1} In this section some useful facts from differential geometry are reminded. Along with the basic vectors $\be_\alpha$ defined in section \ref{sec2}, one can  use vectors of the cobasis $\be^\alpha$. To this end the smooth inverse mapping $\bxi=\bgamma^{-1}(t,\bx)$ is utilized. The cobasic vectors are determined as
\[\be^0=(1,\mathbf{0}),\quad\be^i=(\bxi^i_t,\nabla_{x}\bxi^i),\quad i=1,2,3.\]
Co- and contravariant components of the metric tensor are calculated according to
\[g_{\alpha\beta}=\be_\alpha\cdot\be_\beta,\quad g^{\alpha\beta}=\be^\alpha\cdot\be^\beta.\]
The Christoffel symbols of the second kind are determined by the expansion of derivatives of basic vectors:
\[\pd{\be_\alpha}{t}=\Gamma_{\alpha 0}^\beta\be_\beta,\quad \pd{\be_\alpha}{\xi^i}=\Gamma_{\alpha i}^\beta\be_\beta.\]
Note, that the special form of basic vectors $\be_\alpha$ yields
\[\Gamma_{\alpha\beta}^0=0.\]
The expressions for Christoffel symbols in terms of components of the metric tensor are
\[\Gamma_{\alpha\beta}^\gamma=\frac{1}{2}\,g^{\gamma\delta}\left(\pd{g_{\alpha\delta}}{\xi^\beta}+
\pd{g_{\beta\delta}}{\xi^\alpha}-\pd{g_{\alpha\beta}}{\xi^\delta}\right).\]
The following equality takes place:
\[\Gamma_{\alpha\beta}^\beta=\Gamma_{\alpha\,i}^i=\frac{1}{\sqrt{g}}\pd{\sqrt{g}}{\xi^\alpha},\quad g=\det||g_{\alpha\beta}||.\]
The covariant derivative of a scalar function coincides with partial derivatives with respect to the corresponding coordinate $\xi^\alpha$. The covariant derivative of the contravariant coordinate of a vector field has the form
\[u^i_{,\,\alpha}=\pd{u^i}{\xi^\alpha}+u^\beta\Gamma_{\beta\alpha}^i.\]

\section{Group classification}\label{A2} Here the details of the group classification of system \eqref{MainCompr} with respect to function $h(p,\rho)$ as an ``arbitrary element'' are given. Results of the admissible group calculations are summed up in Theorem \ref{th2}. The classifying equation is \eqref{ClassEq}. One need to specify all possible forms of function $h(p,\rho)$ up to equivalence transformations \eqref{EquivTrans}, and the corresponding restrictions on arbitrary constants $c_i$ such that equation \eqref{ClassEq} is satisfied identically for $p$ and $\rho$.

The classifying parameter is the dimension $\dim\{\bV\}$ of the linear space spanned by the vector
\[\bV=(h+4ph_p+5\rho h_\rho,ph_p+\rho h_\rho,h_p)\]
for various values of $p$ and $\rho$.

a) $\dim\{\bV\}=0.$ This implies
\[h+4ph_p+5\rho h_\rho=0,\quad ph_p+\rho h_\rho=0,\quad h_p=0.\]
Hence, function $h$ is identically zero. Constants $c_1$, $c_2$, $c_3$ take arbitrary values. This case gives the maximal extension of the admissible group.

b) $\dim\{\bV\}=1$. Hence, vector $\bV$ is proportional to the constant vector:
\[h+4ph_p+5\rho h_\rho=m U(p,\rho),\quad ph_p+\rho h_\rho=n U(p,\rho),\quad h_p=s U(p,\rho).\]
The equivalent form of these equations is
\[h=(m - 5 n + p s) U,\quad h_\rho=\rho^{-1}(n - p s) U,\quad h_p=s U.\]
Calculation of mixed derivatives yields the following equations for function $U(p,\rho)$:
\begin{equation}\label{Class1}
\begin{array}{l}
(m - 5 n + p s) U_p=0,\quad \rho^{-1}(n - p s) U=(m - 5 n + p s) U_\rho,\\[2mm]
s U+s \rho U_\rho=(n-ps)U_p.
\end{array}
\end{equation}
First of equations \eqref{Class1} and condition $h\ne0$ give $U_p=0$, hence
\[U=U(\rho).\]
To satisfy the last equation of \eqref{Class1} the following two cases must be separated: $s=0$ and $s\ne 0$.

\underline{$s=0$}. Solution of the remaining second equation in \eqref{Class1} is
\[U=\rho^\frac{n}{m-5n}U_0.\]
Accurate to the equivalence transformations \eqref{EquivTrans} function $h$ can be brought to the form
\[h=\rho^k.\]
Hence, vector $\bV$ is equal to
\[\bV=\rho^k(1+5k,k,0).\]
The classifying equation \eqref{ClassEq} is reduced to
\[c_1(1+5k)-2kc_2=0,\quad c_3 \mbox{ is arbitrary}.\]
Hence, the admissible group is extended by two operators.

\underline{$s\ne0$}. The last of equations \eqref{Class1} gives
\[U=\frac{U_0}{\rho}.\]
From the second of equations \eqref{Class1} it follows that $m=4n$. Accurate to the equivalence, function $h$ can be brought to the form
\[h= k  p/\rho.\]
with an arbitrary constant $k$. Hence,
\[\bV=(0,0,1/\rho)\]
which implies that $c_1$ and $c_2$ are arbitrary constants, and $c_3=0$. The group is extended by two operators.

c) $\dim\{\bV\}=2$. In this case there exists a constant vector $(m,n,s)$ orthogonal to vector $\bV$ for every $\rho$ and $p$, that is
\begin{equation}\label{Class2}
m(h+4ph_p+5\rho h_\rho)+n(ph_p+\rho h_\rho)+sh_p=0.
\end{equation}
Integration of this equation in the case $4m+n\ne 0$ for function $h(p,\rho)$ gives
\[h=\bigl((4m+n)p+s\bigr)^\frac{-m}{4m+n}f\left(\rho\,\bigl((4m+n)p+s\bigr)^{-\frac{5m+n}{4m+n}}\right)\]
with an arbitrary function $f$. Accurate to transformations \eqref{EquivTrans} this is equivalent to
\begin{equation}\label{Class11}
h=p^k f(\rho\,p^{k-1}).
\end{equation}
with some constant parameter $k$. Equation \eqref{ClassEq} is satisfied only for
\[(1+4k)c_1 - 2 c_2 k=0,\quad c_3=0.\]
The admissible group is extended by one operator. In the case $4m+n=0$, $s\ne0$ the general solution of equation \eqref{Class2} has the following form:
\[h=e^{-m p/s}f(\rho\,e^{-mp/s}),\]
which is equivalent to either
\begin{equation}\label{Class12}
h=f(\rho\, e^p)/\rho
\end{equation}
for $m\ne 0$, or
\begin{equation}\label{Class13}
h=f(\rho)
\end{equation}
otherwise. Equation \eqref{ClassEq} gives
\[c_2=c_3=2c_1\]
for the case \eqref{Class12}, and
\[c_1=c_2=0\]
for the case \eqref{Class13}. Finally, the case $4m+n=0$, $s=0$ gives the following solution
\begin{equation}\label{Class14}
h=f(p)/\rho.
\end{equation}
The classifying equation \eqref{ClassEq} yields
\[c_2=2c_1,\quad c_3=0.\]
Hence, class of function $h$ which gives two-dimensional vector space $\{\bV\}$ up to the equivalence is exhausted by representatives \eqref{Class11}--\eqref{Class14}.

d) $\dim\{\bV\}=3$. This case is equivalent to an arbitrary function $h$. The admissible group coincides with the kernel.

\bibliographystyle{unsrt}
\bibliography{Literature}
\end{document}